\input epsf.sty

\documentclass[12pt,a4paper]{article}
\usepackage{amssymb}
\usepackage{amsmath}
\usepackage{graphics}
\usepackage{pst-all}
\usepackage{cite}
\usepackage{hyperref}
\usepackage{indentfirst}
\usepackage{makeidx}
\usepackage[utf8]{inputenc}
\usepackage[T1]{fontenc}
\usepackage{epsfig}
\usepackage{subfigure}
\usepackage{rotating}
\newcommand{\ud}{\,\mathrm{d}}

\begin{document}
\title{Study of the bilinear biquadratic Heisenberg model on a honeycomb lattice via Schwinger bosons}
\author{A. R. Moura}
\begin{center}
{\bf\large Study of the bilinear biquadratic Heisenberg model on a honeycomb lattice via Schwinger bosons}\\
\vspace{0.5cm}
Antônio R. Moura$^\dagger$ and Afrânio R. Pereira$^\ddagger$\\
{\it $^\dagger$ Universidade Federal de Uberlândia, Minas Gerais, Brazil\\
$^\ddagger$ Universidade Federal de Viçosa, Minas Gerais, Brazil.}
\end{center}

\begin{abstract}
We analyze the biquadratic bilinear Heisenberg magnet on a honeycomb lattice
via Schwinger boson formalism. Due to their vulnerability to quantum fluctuations,
non conventional lattices (kagome, triangular and honeycomb for example)
have been cited as candidates to support spin liquid states. Such states without long
range order at zero temperature are known in one-dimensional spin models but their
existence in higher dimensional systems is still under debate. Biquadratic interaction is
responsible for various possibilities and phases as it is well-founded for one-dimensional systems.
Here we have used a bosonic representation to study the properties at zero and finite
low temperatures of the biquadratic term in the two-dimensional hexagonal honeycomb lattice.
The results show a ordered state at zero temperature but much more fragile than
that of a square lattice; the behavior at finite low temperatures is in accordance
with expectations.
\end{abstract}

\section{Introduction}
Non conventional lattices in magnetic systems have received much attention in recent
years. Traditional square lattices are well established and no surprises are expected.
On the other hand, the non conventional lattices are serious candidates
to the so-called two-dimensional spin liquid phase. Spin liquids are disordered states of matter
with power-law decay of spin-spin correlations and zero local magnetic moment. Such properties occur at
zero temperature and the disorder is derived from quantum and not thermal fluctuations.
These phases are known to exist in one-dimensional antiferromagnets
but they still remain uncertain in higher dimensions. Most known two-dimensional ($2D$) magnetic materials
have a N\'{e}el order at zero temperature ($T=0$) even though some unusual systems may be considered as candidates
to present a spin liquid behavior. In three-dimensional magnetic systems, the existence of this state is
even more unexpected due to stronger spin interactions.

One possible way to obtain a $2D$ spin liquid is through the presence of geometric frustration
in some lattices. In the classical antiferromagnetic kagomé lattice, for example, it is
impossible to align all neighbors spins and the ground state is highly degenerate.
In addition, anisotropies together with longer range interactions (second and far nearest neighbors)
contribute to disorder the ground state, increasing the possibility of a spin liquid state.
Although the properties at zero temperature are questionable, recent works indicate
the occurrence of spin
liquid states \cite{JAP69, PRB45, PRB72, PRB82, PRB84, PRL98a, PRL98b, science332, arxiv1012}.
Even lattices without frustration have shown interesting possibilities as it is the case of hexagonal honeycomb lattice.
The honeycomb is the two-dimensional lattice with the smallest coordination number
(neighbors number) $z=3$. It is between the disordered one dimensional spin model
with $z=2$ and the ordered (at zero temperature) square lattice with $z=4$.
Thus, such a system may have larger vulnerability to quantum fluctuations, mainly for small
spins, and it is a possible candidate to be a two-dimensional spin liquid. Nevertheless, recent
works have shown a N\'{e}el order for the spin-$1/2$ Heisenberg AF on honeycomb
lattice \cite{JPCM1, PRB45}, although weaker than the square lattice case. Furthermore, frustrated honeycomb models have revealed disordered ground states \cite{PRB43, PRB49, EPVB20, PRB72-2, PRB74}.

In the present work we are interested in the behavior of the biquadratic bilinear Heisenberg
model in a hexagonal honeycomb lattice. It is well known that the Heisenberg Hamiltonian is one of
the simplest model able to describe both the ferromagnetic and antiferromagnetic materials in any dimension.
Here, in addition to the usual Heisenberg model, we consider a biquadratic term in such a way that the complete hamiltonian is given by:
\begin{equation}
\label{hamiltonian_bb}
H=\sum_{\langle i,j\rangle}\left[ J_1(\vec{S}_i\cdot\vec{S}_j)+J_2(\vec{S}_i\cdot\vec{S}_j)^2\right],
\end{equation}
where the sum is over nearest neighbors and the constants $J_1$ and $J_2$ define
the bilinear and biquadratic couplings, respectively. It is important to highlight
that we adopt $S=1$ once the biquadratic term makes sense only for $S>1/2$. The $SU(2)$
spin rotation symmetry in Hamiltonian (\ref{hamiltonian_bb}) is preserved and so we
can expect Goldstone modes as lower energy excitations over the ground state. We consider
the case $J_1=1$ (antiferromagnetic coupling) and $-J_1\leq J_2\leq J_1$ but it is also common to
write $J_1=\text{cos}\theta$ and $J_2=\text{sin}\theta$. The one-dimensional case is well
documented \cite{PRB44, PRB51, PRB58} and the various phases were already discovered. For
$\theta=\pi$ one has the usual ferromagnet, while $\theta=0$ corresponds to the usual Heisenberg
antiferromagnetic system. In the range $\pi<\theta<5\pi/4$, there is a stable ferromagnetic
regime with long range order (LRO); for $-\pi/4<\theta<\pi/4$ there is an antiferromagnetic
phase with Haldane gap (spin-$1$) and, in the limit $\pi/4<\theta<\pi/2$,
there is a trimerized phase. Some points have exact solutions. For instance,
the angles $\theta=\pm\pi/4$ can be solved by the Bethe ansatz and
besides, the angle $\tan\theta=1/3$ corresponds to AKLT Model.
The two-dimensional case is more complicated and only some regions are known. Ivanov \textit{et al.}
\cite{PRB68, PRB77} have shown a nematic phase for $\theta\gtrsim 5\pi/4$ by using a continuous model
similar to the $O(3)$ nonlinear Sigma Model (the same result has been achieved by
Chubukov using the Holstein-Primakoff representation \cite{PRB43}); for
$5\pi/5<\theta<7\pi/4$ there is a dimerized ferromagnetic phase.

We can represent the biquadratic term as a function of spin and quadrupole operators
(second-order spin moment):
\begin{equation}
\left({\bf S}_i\cdot{\bf S}_j\right)^2=\frac{1}{2}\left({\bf Q}_i\cdot{\bf Q}_j\right)-\frac{1}{2}\left({\bf S}_i\cdot{\bf S}_j\right)+\frac{4}{3},
\end{equation}
where $Q_i$ operators are given by:
\begin{align}
Q_i^{(0)}&=\frac{2(S_i^z)^2-(S_i^x)^2-(S_i^y)^2}{\sqrt{3}},\\
Q_i^{(2)}&=(S_i^x)^2-(S_i^y)^2,\\
Q_i^{xy}&=S_i^x S_i^y+S_i^y S_i^x,\\
Q_i^{yz}&=S_i^y S_i^z+S_i^y S_i^z,\\
Q_i^{zx}&=S_i^z S_i^x+S_i^x S_i^z.
\end{align}
The three spin operators together the five quadrupole operators form the generators
of the $SU(3)$ Lie group. For the special case $J_1=J_2$ one has the $SU(3)$ symmetric
ferromagnetic model \cite{PRB65} while for $J_1=0$ one has the
$SU(3)$ symmetric valence-bound antiferromagnet \cite{NPB265, PRB40, EL9}.

The usual methods used to study systems in condensed matter physics are vast
and diversified. In the current work we have adopted the Schwinger boson representation
to develop the physics of the bilinear biquadratic Heisenberg model in a honeycomb
lattice at both zero and low temperatures. The Schwinger formalism
has some advantages over other bosonic representations (such as Holstein-Primakoff
and Dyson-Maleev). Firstly, the holonomic constraint that fix the
number of bosons can be implemented easily by a Lagrange multiplier.
Secondly, there are not root terms and so we do not need to specify a preferential
direction to the ground state as occur for the Holstein-Primakoff method. Therefore we are able to treat
ordered and disordered phases, which are important in the search for a possible spin liquid state.
Following the usual prospects, we have adopted the boson condensation at zero temperature
to avoid the divergences of the theory and at low temperatures we have used approximations
that, within the correct limits, provide coherent results. The paper is organized as follow:
in Sec. \ref{formalism} we developed the Schwinger bosons formalism; in Sec. \ref{results},
we present the results for zero temperature and finite low temperature and, finally,
the conclusions are exposed in the last section (\ref{conclusions}).

\section{Formalism}
\label{formalism}
Commonly, the spin operators are defined by two $SU(2)$ Schwinger operators but, because
the biquadratic term, we have considered the representation by $SU(3)$ Schwinger formalism.
Thus, each spin operator is represented by three bosonic operators $a_{i,m}$, where $i$
denotes the lattice sites and $m=-1,\ 0,\ 1$ specifies the eigenvalues of $S_i^z$. Accordingly,
$a^\dagger_{i,m}|0\rangle$ creates a particle with $z$-component of spin ($S^z=m$) on
site $i$ (we denote the $|0\rangle$ state as the vacuum of the Fock space). The generators
$F^{mn}_i=a^\dagger_{i,m}a_{i,n}$ form the $SU(3)$ Lie algebra and obey the commutation
relation $[F^{mn}_i,F^{pq}_j]=\delta_{i,j}(\delta_{n,p}F^{mq}_i-\delta_{m,p}F^{nq}_i)$.
As a function of the $a$ operators, the spin operators are expressed by $S_i^+=\sqrt{2}
(a_{i,0}^{\dagger}a_{i,-1}+a_{i,1}^{\dagger}a_{i,0})$,
$S_i^-=\sqrt{2}(a_{i,-1}^{\dagger}a_{i,0}+a_{i,0}^{\dagger}a_{i,1})$ and
$S_i^z=a_{i,1}^{\dagger}a_{i,1}-a_{i,-1}^{\dagger}a_{i,-1}$.
The bosonic operators keep the spin commutation relation and to fix $S^2_i=S(S+1)$ we
have to impose the local constraint $\sum_m a^\dagger_{i,m}a_{i,m}=S$. In order to
symmetrize the spin and quadrupole operators, we apply a rotation over the $a$ operators
defining new operators $b$ as follow:
\begin{align}
b_{i1}&=\frac{1}{\sqrt{2}}\left(a_{i,-1}-a_{i,1}\right),\\
b_{i2}&=\frac{-i}{\sqrt{2}}\left(a_{i,-1}+a_{i,1}\right),\\
b_{i3}&=a_{i,0}.
\end{align}
Therefore, the spin operators are written as:
\begin{align}
S_i^x&=-i\left(b_{i2}^\dagger b_{i3}-b_{i3}^{\dagger}b_{i2}\right),\\
S_i^y&=-i\left(b_{i3}^\dagger b_{i1}-b_{i1}^{\dagger}b_{i3}\right),\\
S_i^z&=-i\left(b_{i1}^\dagger b_{i2}-b_{i2}^{\dagger}b_{i1}\right),
\end{align}
and the quadrupoles:
\begin{align}
Q_i^{(0)}&=\frac{1}{\sqrt{3}}\left(b_{i1}^{\dagger}b_{i1}+b_{i2}^{\dagger}b_{i2}-2b_{i3}^{\dagger}b_{i3}\right),\\
Q_i^{(2)}&=-\left(b_{i1}^{\dagger}b_{i1}-b_{i2}^{\dagger}b_{i2}\right),\\
Q_i^{xy}&=-\left(b_{i1}^{\dagger}b_{i2}+b_{i2}^{\dagger}b_{i1}\right),\\
Q_i^{yz}&=-\left(b_{i2}^{\dagger}b_{i3}+b_{i3}^{\dagger}b_{i2}\right),\\
Q_i^{zx}&=-\left(b_{i3}^{\dagger}b_{i1}+b_{i1}^{\dagger}b_{i3}\right),
\end{align}
while the constraint holds the same. The biquadratic bilinear Heisenberg Hamiltonian
as a function of $b$ operators is expressed by:
\begin{equation}
H=\sum_{\langle i,j\rangle}\left[(J_2-J_1)\mathcal{A}_{ij}^{\dagger}\mathcal{A}_{ij}+J_1:\mathcal{B}_{ij}^{\dagger}\mathcal{B}_{ij}:\right],
\end{equation}
where we have introduced the bond operators $\mathcal{A}_{ij=}\sum_\mu b_{i\mu}b_{j\mu}$ and
$\mathcal{B}_{ij}=\sum_\mu b_{i\mu}^{\dagger}b_{j\mu}$;
the two points denote normal ordering and $\mu=1, 2, 3$ (distinct of the $m=-1, 0, 1$ index).
The Hamiltonian is fourth order in $b$ and
we decouple it by the Hubbard-Stratonovich transform \cite{SPD2,PRL3}:
\begin{equation}
\Xi^\dagger_{ij}\Xi_{ij}\rightarrow\langle\Xi^\dagger_{ij}\rangle\Xi_{ij}+
\langle\Xi_{ij}\rangle\Xi^\dagger_{ij}-\langle\Xi^\dagger_{ij}\rangle\langle\Xi_{ij}\rangle.
\end{equation}
In above equation $\Xi_{ij}= \mathcal{A}_{ij},\mathcal{B}_{ij}$, where we have adopted the
mean field $A=\langle\mathcal{A}^\dagger_{ij}\rangle=\langle\mathcal{A}_{ij}\rangle$ and
$B=\langle B^\dagger_{ij}\rangle=\langle\mathcal{B}_{ij}\rangle$.
Therefore the second order mean field Hamiltonian is given by:
\begin{align}
H^{\textrm{MF}}&=-\frac{3N}{2}\left[(J_2-J_1)A^2+J_1B^2\right]-NS\lambda+\lambda\sum_{i\mu}b_{i\mu}^{\dagger}b_{i\mu}+\nonumber\\
&+\sum_{\langle i,j\rangle}\left[(J_2-J_1)A\left(\mathcal{A}_{ij}^{\dagger}+\mathcal{A}_{ij}\right)+J_1B\left(\mathcal{B}_{ij}^{\dagger}+\mathcal{B}_{ij}\right)\right].
\end{align}
The constraint $\sum_\mu b^\dagger_{i\mu}b_{i\mu}=S$ is implemented by a Lagrange multiplier
$\lambda_i$ on each site and we have adopted a constant mean field value
$\lambda=\langle\lambda_i\rangle$. The $\lambda$ parameter is similar to the chemical potential $\mu$
while the constraint $\sum_\mu b^\dagger_{i\mu}b_{i\mu}=S$ counts the
bosons number on each site. The mean field values $A$, $B$ and $\lambda$ are determined by
minimizing the Helmholtz free energy.

The honeycomb is a bipartite but not a Bravais lattice and so we have to treat each
sublattice separately. The sublattices $R$ and $R^\prime$ are hexagonal Bravais lattices and
they are coupled by nearest neighbors interactions. After Fourier transforming the
Schwinger bosons independently on each sublattice, we obtain
\begin{equation}
b_{i\mu}=\sqrt{\frac{2}{N}}\sum_{\bf k}  e^{i{\bf k} \cdot{\bf r}_i}b_{{\bf k} \mu}, \quad i\in R
\end{equation}
and:
\begin{equation}
b_{j\mu}=\sqrt{\frac{2}{N}}\sum_{\bf k}  e^{i{\bf k} \cdot{\bf r}_j}b_{{\bf k} \mu}^\prime, \quad j\in R^\prime,
\end{equation}
the Hamiltonian is written as:
\begin{align}
\label{hamiltonian_k}
H^{\textrm{MF}}&=H_0+\frac{1}{2}\sum_{\bf k}\sum_\mu \left[3(J_2-J_1)A\left(b_{{\bf k} \mu}^{\dagger}b_{-{\bf k} \mu}^{\prime\dagger}\gamma_{\bf k} +b_{{\bf k} \mu}^{\prime\dagger}b_{-{\bf k} \mu}^{\dagger}\gamma_{\bf k} ^{\ast}\right)+\right.\nonumber\\
&\left.+3J_1B\left(b_{{\bf k} \mu}^{\dagger}b_{{\bf k} \mu}^\prime\gamma_{\bf k} +b_{{\bf k} \mu}^{\prime\dagger}b_{{\bf k} \mu}\gamma_{\bf k} ^{\ast}\right)+\lambda\left(b_{{\bf k} \mu}^{\dagger}b_{{\bf k} \mu}+b_{{\bf k} \mu}^{\prime\dagger}b_{{\bf k} \mu}^\prime\right)+h.c.\right],
\end{align}
where $H_0=-\frac{3N}{2}\left[(J_2-J_1)A^2+J_1B^2\right]-NS\lambda$ are constant terms and $\gamma_{\bf k}=e^{i\varphi_{\bf k}}|\gamma_{\bf k}|$
is the structure factor:
\begin{align}
\label{structure}
\gamma_{\bf k}=\frac{1}{3}\left[2\textrm{cos}\frac{k_x}{2}\textrm{cos}\frac{\sqrt{3}k_y}{2}+\textrm{cos}k_x+2i\textrm{sin}\frac{k_x}{2}\textrm{cos}\frac{\sqrt{3}k_y}{2}-i\textrm{sin}k_x\right].
\end{align}

In Hamiltonian (\ref{hamiltonian_k}), the two sublattices are still coupled and
differently from the square lattice, the structure factor for honeycomb lattice is not real.
We solve these two difficulties defining new operators
$b_{{\bf k} \mu}=\frac{e^{i\varphi_{\bf k}/2}}{\sqrt{2}}\left(ic_{{\bf k} \mu}^I+c_{{\bf k} \mu}^{II}\right)$
and $b_{{\bf k} \mu}^\prime=\frac{e^{-i\varphi_{\bf k}/2}}{\sqrt{2}}\left(-ic_{{\bf k} \mu}^I+c_{{\bf k} \mu}^{II}\right)$. The new bosons $c_{\bf k}$ satisfy all commutation relations and the Hamiltonian is written as:
\begin{equation}
\label{hamiltonian_c}
H^{\textrm{MF}}=H_0+\frac{1}{2}\sum_{\bf k} {\bf \beta}_{\bf k} ^{I\dagger}\tilde{H}^I{\bf \beta}_{\bf k} ^I+\frac{1}{2}\sum_{\bf k} {\bf \beta}_{\bf k} ^{II\dagger}\tilde{H}^{II}{\bf \beta}_{\bf k} ^{II},
\end{equation}
with ${\bf \beta}_{\bf k} ^{s\dagger}=(c_{{\bf k} 1}^{s\dagger},c_{{\bf k} 2}^{s\dagger},c_{{\bf k} 3}^{s\dagger},c_{-{\bf k} 1}^s,c_{-{\bf k} 2}^s,c_{-{\bf k} 3}^s)$
where $s=I,II$, while the matrices are $\tilde{H}^I=(\lambda-3J_1B|\gamma_{\bf k}|)\sigma_0\otimes I_{3\times3}+3(J_2-J_1)A|\gamma_{\bf k}|\sigma_x\otimes I_{3\times3}$ and
$\tilde{H}^{II}=(\lambda+3J_1B|\gamma_{\bf k}|)\sigma_0\otimes I_{3\times3}+3(J_2-J_1)A|\gamma_{\bf k}|\sigma_x\otimes I_{3\times3}$ (here $\sigma_i$ are the Pauli matrices). $H^{\textrm{MF}}$ can be diagonalized by a canonical Bogoliubov transformation:
\begin{align}
\label{coperator}
c_{{\bf k} \mu}^{I}&=\textrm{cosh}\theta_{\bf k}^{I}\alpha_{{\bf k} \mu}^{I}+\textrm{sinh}\theta_{\bf k}^{I} \alpha_{-{\bf k } \mu}^{I\dagger}\\
\label{coperator2}
c_{{\bf k} \mu}^{II}&=\textrm{cosh}\theta_{\bf k}^{II}\alpha_{{\bf k} \mu}^{II}+\textrm{sinh}\theta_{\bf k}^{II} \alpha_{-{\bf k} \mu}^{II\dagger}.
\end{align}
We choose $\theta_{\bf k}^I$ and $\theta_{\bf k}^{II}$ so that the non-diagonal terms vanish.
It is achieved by:
\begin{align}
\textrm{tanh}2\theta_{\bf k}^I&=-\frac{\lambda-3J_1B|\gamma_{\bf k} |}{3(J_2-J_1)A|\gamma_{\bf k} |},\\
\textrm{tanh}2\theta_{\bf k}^{II}&=-\frac{\lambda+3J_1B|\gamma_{\bf k} |}{3(J_2-J_1)A|\gamma_{\bf k} |}.
\end{align}
Once diagonalized, $H^{\textrm{MF}}$ gives the eigenvalues of energy:
\begin{align}
\label{energy_I}
E_I&=\sqrt{\left(\lambda-3J_1B|\gamma_{\bf k} |\right)^2-\left(3A(J_2-J_1)|\gamma_{\bf k}|\right)^2}\\
\label{energy_II}
E_{II}&=\sqrt{\left(\lambda+3J_1B|\gamma_{\bf k} |\right)^2-\left(3A(J_2-J_1)|\gamma_{\bf k}|\right)^2}.
\end{align}
The mean field equations are $\partial F/\partial A$, $\partial F/\partial B$ and
$\partial F/\partial\lambda$, where the Helmholtz free energy $F$ is
\begin{equation}
\label{free_energy}
F=H_0+\frac{3}{\beta}\sum_{\bf k} \left\{\ln\left[\textrm{sinh}\left(\frac{\beta E_I}{2}\right)\right]+\ln\left[\textrm{sinh}\left(\frac{\beta E_{II}}{2}\right)\right]\right\},
\end{equation}
yielding the integral self-consistent equations:
\begin{equation}
\label{escSft}
S+\frac{3}{2}=\frac{3}{2N}\sum_{\bf k} \left[\textrm{coth}\left(\frac{\beta E_I}{2}\right)\frac{\lambda-3J_1B|\gamma_{\bf k} |}{E_I}+\textrm{coth}\left(\frac{\beta E_{II}}{2}\right)\frac{\lambda+3J_1B|\gamma_{\bf k} |}{E_{II}}\right],
\end{equation}
\begin{equation}
\label{escAft}
A=-\frac{3}{2N}\sum_{\bf k} \left[\textrm{coth}\left(\frac{\beta E_I}{2}\right)\frac{3A(J_2-J_1)}{E_I}+\textrm{coth}\left(\frac{\beta E_{II}}{2}\right)\frac{3A(J_2-J_1)}{E_{II}}\right]|\gamma_{\bf k} |^2
\end{equation}
and
\begin{equation}
\label{escBft}
B=\frac{3}{2N}\sum_{\bf k} \left[\textrm{coth}\left(\frac{\beta E_{II}}{2}\right)\frac{\lambda+3J_1B|\gamma_{\bf k}|}{E_{II}}-\textrm{coth}\left(\frac{\beta E_I}{2}\right)\frac{\lambda-3J_1B|\gamma_{\bf k}|}{E_I}\right]|\gamma_{\bf k}|.
\end{equation}
As it is well known, one- and two-dimensional systems can have LRO only at zero temperature and
this implies an abrupt change at $T=0$. Indeed, when the temperature approaches absolute zero, the bosons
condensate at a zero energy state and the self-consistent equations diverge. This
inconvenience is surmounted by separating the divergent term of the sum and introducing a
new parameter (the condensate density) as it is done in the Bose-Einstein condensate.
For finite temperatures such problem does not exist and the equations can be solved
customarily. Obviously, the self-consistent equation can not be solved exactly and numeric
methods or approximations are usually applied. In the next section we will present the results for both
$T=0$ and low finite temperatures.

Using equations (\ref{coperator}) and (\ref{coperator2}), we calculate the mean
field double boson operators:
\begin{align}
\label{mfs}
\langle b_{i\mu}^\dagger b_{j\mu}\rangle &=\frac{1}{N}\sum_{\bf k}e^{-i{\bf k}\Delta {\bf r}}\left[\textrm{cosh}2\theta_{\bf k}^I\left(n_{\bf k}^I+\frac{1}{2}\right)+\textrm{cosh}2\theta_{\bf k}^{II}\left(n_{\bf k}^{II}+\frac{1}{2}\right)\right]-\frac{1}{2}\delta_{ij}\\
\label{mfs2}
\langle b_{i\mu}b_{j\mu}\rangle &=\frac{1}{N}\sum_{\bf k}e^{-i{\bf k}\Delta {\bf r}}\left[-\textrm{sinh}2\theta_{\bf k}^I\left(n_{\bf k}^I+\frac{1}{2}\right)+\textrm{sinh}2\theta_{\bf k}^{II}\left(n_{\bf k}^{II}+\frac{1}{2}\right)\right]
\end{align}
for $i$ and $j$ belonging to the same sublattice and:
\begin{align}
\label{mfd}
\langle b_{i\mu}^\dagger b_{j\mu}\rangle &=\frac{1}{N}\sum_{\bf k}e^{-i{\bf k}\Delta {\bf r}}e^{-i\varphi_{\bf k}}\left[-\textrm{cosh}2\theta_{\bf k}^I\left(n_{\bf k}^I+\frac{1}{2}\right)+\textrm{cosh}2\theta_{\bf k}^{II}\left(n_{\bf k}^{II}+\frac{1}{2}\right)\right]\\
\langle b_{i\mu}b_{j\mu}\rangle &=\frac{1}{N}\sum_{\bf k}e^{-i{\bf k}\Delta {\bf r}}e^{-i\varphi_{\bf k}}\left[\textrm{sinh}2\theta_{\bf k}^I\left(n_{\bf k}^I+\frac{1}{2}\right)+\textrm{sinh}2\theta_{\bf k}^{II}\left(n_{\bf k}^{II}+\frac{1}{2}\right)\right]
\end{align}
for $i$ and $j$ of different sublattices. The phase angle $\varphi_{\bf k}$ is the same as that which
appears in the structure factor (equation (\ref{structure})), whilst the boson densities are given by:
\begin{equation}
n_{\bf k}^t=\langle \alpha_{{\bf k}\mu}^{t\dagger}\alpha_{{\bf k}\mu}^{t}\rangle=\frac{1}{e^{\beta E_t}-1}
\end{equation}
with $t=I,\ II$. All others mean field quantities are null.

\section{Results}
\label{results}
\begin{figure}[!h]
	\centering
	\includegraphics[scale=1]{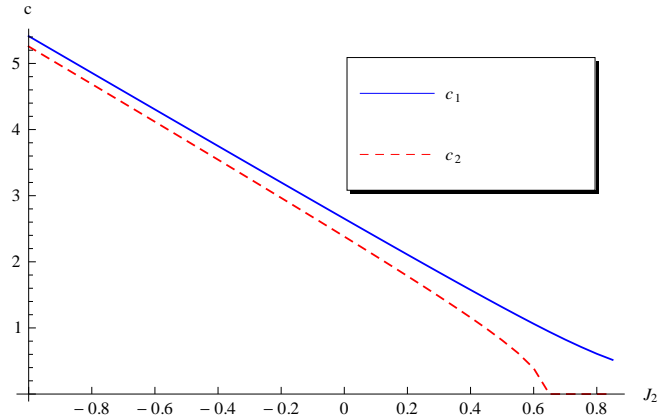}
	\caption{Almost linear decreasing behavior of spin-wave velocities $c_1$ and $c_2$.}
	\label{spinwave}
\end{figure}

When the temperature decreases to zero, a phase transition takes place and one of the
energies of the spectrum vanishes, characterizing a boson condensation.
Therefore, the self-consistent equations diverge and there are not more solutions
for the parameters $A$, $B$ and $\lambda$. According to Takahashi and Arovas \textit{et al.}
\cite{PRL58, PRB38, PRB40b, PRB40c}, the non-existence of
solutions is related to a spontaneous broken symmetry, since Schwinger formalism is
invariant over $SU(2)$. At finite temperatures, there are solutions for any dimension and the
system is disordered, i.e., there is not long range order (at zero temperature, there are solutions only
for the one-dimensional case). Here, the condensation occurs for
$E_I$ if $J_1>0$ and $E_{II}$ otherwise. The ground state is therefore ordered and the
lowest excitation energies are the massless Goldstone modes. Expanding for small $k$,
the dispersion relations (\ref{energy_I}) and (\ref{energy_II}) assume a relativist form:
\begin{equation}
E_I=\sqrt{\Delta_{I}^2+{\bf k} ^2c_{I}^2}\quad\textrm{and}\quad E_{II}=\sqrt{\Delta_{II}^2+{\bf k} ^2c_{II}^2},
\end{equation}
with the gap energies:
\begin{align}
\Delta_{I}&=\sqrt{(\lambda-3J_1B)^2-(3(J_1-J_2)A)^2},\\
\Delta_{II}&=\sqrt{(\lambda+3J_1B)^2-(3(J_1-J_2)A)^2},
\end{align}
while the spin-wave velocities are given by:
\begin{align}
c_{1}&=\sqrt{\frac{1}{2}\left(3J_1B-9J_1^2B^2+9(J_1-J_2)^2A^2\right)},\\
c_{2}&=\sqrt{\frac{1}{2}\left(-3J_1B-9J_1^2B^2+9(J_1-J_2)^2A^2\right)}.
\end{align}
The spin-wave velocities as a function of $J_2$ ($J_1=1$) are plotted in Fig. (\ref{spinwave}).
A similar behavior is observed for the frustrated honeycomb Heisenberg system\cite{PRB49},
where there is a linear decreasing of the spin-wave velocity as a function of the
second-nearest neighbors exchange coupling. For $J_2\geq0.65$, the $c_2$ spin-wave
velocity is null whilst the point where $c_1=0$ is beyond of the limits considered.

Considering positive values of $J_1$, the condensation occurs for $E_{I}$ and
then $\Delta_I=0$ while $\Delta_{II}$ is finite. After separating the
divergent terms, the self-consistent equations are written in the continuous limit as:
\begin{equation}
\label{escSzt}
\rho=\left(S+\frac{3}{2}\right)-\frac{3}{2}\int\frac{\ud^2{\bf k}}{2\sigma} \left[\frac{\lambda-3J_1B|\gamma_{\bf k} |}{E_I}+\frac{\lambda+3J_1B|\gamma_{\bf k} |}{E_{II}}\right],
\end{equation}
\begin{equation}
\label{escAzt}
A=\rho-\frac{3}{2}\int\frac{\ud^2{\bf k}}{2\sigma} \left[\frac{3A(J_2-J_1)|\gamma_{\bf k} |^2}{E_I}+\frac{3A(J_2-J_1)|\gamma_{\bf k} |^2}{E_{II}}\right]
\end{equation}
and
\begin{equation}
\label{escBzt}
B=-\rho+\frac{3}{2}\int\frac{\ud^2{\bf k}}{2\sigma} \left[-\frac{(\lambda-3J_1B|\gamma_{\bf k} |)|\gamma_{\bf k} |}{E_I}+\frac{(\lambda+3J_1B|\gamma_{\bf k} |)|\gamma_{\bf k} |}{E_{II}}\right],
\end{equation}
where $\rho$ is a new parameter that measures the boson condensate and
$\sigma=\frac{8\pi}{3\sqrt{3}}$ is the first Brillouin zone area.

\begin{figure}[!h]
	\centering
	\includegraphics[scale=1]{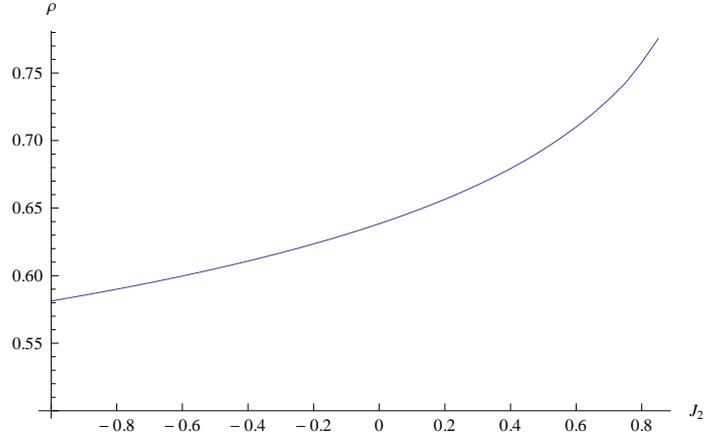}
	\caption{The boson density condensate as a function of $J_2$.}
	\label{density}	
\end{figure}

The condensate density is plotted in figure \ref{density}. For $J_2=0$ we have the pure
Heisenberg antiferromagnet, in which the condensate density is approximately $0.64$.
As expected, $\rho$ is smaller than the corresponding one for the square lattice
(approximately 80$\%$ smaller \cite{PRB40c}). The condensate density
increases as $J_2$ increases (with an almost linear behavior in the range
$-1\leq J_2 \leq 0.2$). Comparing with the spin-wave velocity graphics in Fig.(\ref{spinwave})
we can see that the increasing condensate density (and consequently
the ordering) occurs together with a decreasing of the spin-wave velocities. This is
expected since the spin-wave is responsible for disordering the ground state and then,
the higher velocity implies in higher disorder. Therefore the honeycomb lattice holds a
long range order at $T=0$ but due to the small coordination number this ordering is weaker
than that of the square lattice case.

\begin{figure}[!h]
	\centering
	\includegraphics[scale=1]{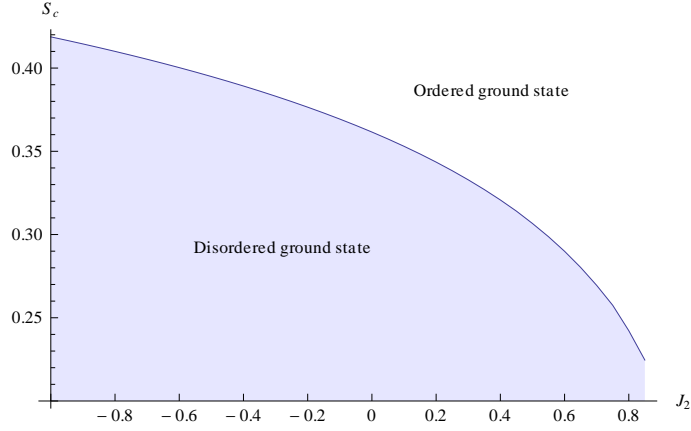}
	\caption{The critical spin value separates the region with an ordered ground
	state from that with a disordered ground state.}
	\label{sc}
\end{figure}

As it is well known, the two-dimensional square lattice presents LRO for all
spin values $S>S_c\approx0.19$. Writing the magnetization as $m_s=S+\frac{3}{2}-\frac{3}{2}\int
\left(\frac{\lambda-3J_1B|\gamma_{\bf k} |}{E_I}+\frac{\lambda+3J_1B|\gamma_{\bf k}|}
{E_{II}}\right)\frac{\ud^2{\bf k}}{2\sigma}$ we determine the critical value of the spin $S_c$
for which $m_s\rightarrow 0$. The results are shown in Fig. \ref{sc}. The $S_c$ curve separates
the region with an ordered ground state from the disordered ground state. As well as
to the square lattice, on the honeycomb lattice the disordered ground state is
inaccessible to all physical spins.
Using equations (\ref{mfs}) and {\ref{mfs2}) we obtain the mean value
$\langle{\bf S}_i\cdot{\bf S}_i\rangle\approx2.5$ for all values of $J_2$. It is
greater than the expected value $S(S+1)=2$ by a factor of approximately $3/2$. Such factor also
appears in the equations obtained by Arovas and Auerbach \cite{PRB38}
and this discrepancy arises because we imposed the constraint only on the average.
The problem can be solved through a
pertubative expansion in low order. The Fourier transform $S(\bf q)$ of the two-point function $\langle{\bf S}_i\cdot{\bf S}_j\rangle$
is as follow:
\begin{align}
S({\bf q})&=-\frac{6}{4}+\frac{3}{N}\sum_{\bf k}\left\{\left[\cosh(2\theta_{\bf k+q}^I-2\theta_{\bf k}^I)\left(n_{\bf k+q}^I+\frac{1}{2}\right)\left(n_{\bf k}^I+\frac{1}{2}\right)+\right.\right.\nonumber\\
&\left.+\cosh(2\theta_{\bf k+q}^{II}-2\theta_{\bf k}^{II})\left(n_{\bf k+q}^{II}+\frac{1}{2}\right)\left(n_{\bf k}^{II}+\frac{1}{2}\right)\right]\left[1+e^{\varphi_{\bf k+q}-\varphi_{\bf k}}\right]+\nonumber\\
&+\left[1-e^{\varphi_{\bf k+q}-\varphi_{\bf k}}\right]\left[\cosh(2\theta_{\bf k+q}^I+2\theta_{\bf k}^{II})\left(n_{\bf k+q}^I+\frac{1}{2}\right)\left(n_{\bf k}^{II}+\frac{1}{2}\right)+\right.\nonumber\\
&\left.\left.+\cosh(2\theta_{\bf k+q}^{II}+2\theta_{\bf k}^{I})\left(n_{\bf k+q}^{II}+\frac{1}{2}\right)\left(n_{\bf k}^{I}+\frac{1}{2}\right)\right]\right\}.
\end{align}
The static uniform susceptibility $\chi=\frac{S({\bf 0})}{3T}$ is therefore:
\begin{equation}
\chi=T^{-1}\frac{2}{N}\sum_{\bf k}\left[n_{\bf k}^I(n_{\bf k}^I+1)+n_{\bf k}^{II}(n_{\bf k}^{II}+1)\right].
\end{equation}
Similar equations were obtained by Takahashi \cite{PRB40b} for an antiferromagnetic system in a
square lattice.

For $T\neq0$ the energies (\ref{energy_I}) and (\ref{energy_II}) are finite and we have
no divergences in the self-consistent equations. The equations (\ref{escSft}),
(\ref{escAft}) and (\ref{escBft}), therefore, can be solved using numeric
methods but we have adopted an approximation following Yoshida \cite{JPSJ58}.
Matching equations (\ref{escSft}) and (\ref{escSzt}) we got:
\begin{align}
\label{densityft}
\rho&=\frac{3}{2}\int\frac{\ud^2{\bf k} }{2\sigma}\left[\frac{\lambda-3J_1B|\gamma_{\bf k} |}{E_I}\textrm{coth}\left(\frac{\beta E_I}{2}\right)+
\frac{\lambda+3J_1B|\gamma_{\bf k} |}{E_{II}}\textrm{coth}\left(\frac{\beta E_{II}}{2}\right)\right]-\nonumber\\
&-\frac{3}{2}\int\frac{\ud^2{\bf k} }{2\sigma}\left[\frac{\lambda_0-3J_1B_0|\gamma_{\bf k} |}{E_{0,I}}+\frac{\lambda_0-3J_1B_0|\gamma_{\bf k} |}{E_{0,II}}\right],
\end{align}
where the "0" index indicates the solutions at zero temperature. The above equation
is solved in the low temperature limit. We separate the integral in two regions:
the first one is a circle around the origin of radius $k_M$ and the other is the
remaining area of the Brioullin zone. The radius is chosen such that the thermal energy is
much lower than the spin-wave energy, i.e. $T\ll c k_M$ ($k_B=1$) where
the spin-wave velocity $c$ refers to $c_1$ or $c_2$. In principle, the superior
value for $k_M$ is much smaller than $1$ and it should be chosen so that the energies can be approximated
by $E_I=\sqrt{\Delta_I^2+{\bf k} ^2c_I^2}$ and $E_{II}=\sqrt{\Delta_{II}^2+{\bf k} ^2c_{II}^2}$.
For $k_M=1$ the error between the exact and approximate energy is around $10\%$, which
allow us to assign $k_M\sim 0.1$ as a reasonable limit.
Thus, we can estimate $k_M$ for temperatures not too small.
At zero temperature, we have $E_{0,I}=kc_{0,I}$ (massless gap mode) and
$E_{0,II}=\sqrt{\Delta_{0,II}^2+{\bf k} ^2c_{0,II}^2}$. Inside the first region
$\frac{\beta E_{II}}{2}\approx\frac{\beta\Delta_{II}^2}{2}\gg1$ and so
$\textrm{coth}\left(\frac{\beta E_{II}}{2}\right)\approx 1$ whilst $E_{I}$ is gapped and
$\textrm{coth}\left(\frac{\beta E_I}{2}\right)$  holds without more approximations.
In the extern region, $k>k_M$, the spin-wave energies are not too small to be approached by the
relativistic dispersion relation; meanwhile the cotangent terms (at low temperatures)
are taken as unitary. With these considerations and after some work, the density
condensate is calculated as:
\begin{align}
\rho&=\frac{6\pi(\lambda-3J_1B)T}{\sigma c_I^2}\left[\ln\textrm{sinh}\left(\frac{\sqrt{\Delta_I^2+{\bf k} _M^2c_I^2}}{2T}\right)-\ln\textrm{sinh}\left(\frac{\Delta_I}{2T}\right)\right]-\nonumber\\
&-\frac{3\pi(\lambda_0-3J_1B_0){\bf k} _M}{\sigma c_{0,I}},
\end{align}
which, at low temperature limit, yields $\Delta_I(T)=Te^{-\kappa T}$ with
$\kappa=\frac{\rho\sigma c_I^2}{6\pi(\lambda-3J_1B)}$. In Fig. (\ref{gap}) we show
the gap energy $\Delta_I$ as a function of the temperature for three different exchange
constants $J_2$ and in Fig. (\ref{lngap}), $\ln\Delta_I$ is showed as a
function of $J_2$ for $T=0.15$. The finite gap result agrees with the Mermin-Wagner theorem.
\begin{figure}[!h]
	\centering
	\includegraphics[scale=1]{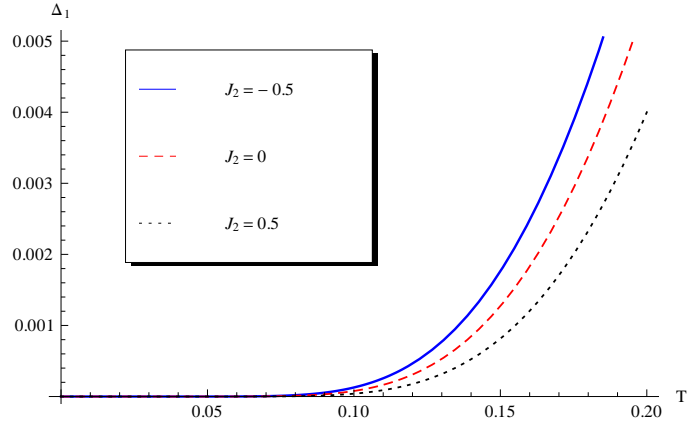}
	\caption{The gap energy $\Delta_I$ as a function of temperature for some values of $J_2$ ($\Delta_{II}=0$ in the considered limits).}
	\label{gap}
\end{figure}

\begin{figure}[!h]
	\centering
	\includegraphics[scale=1]{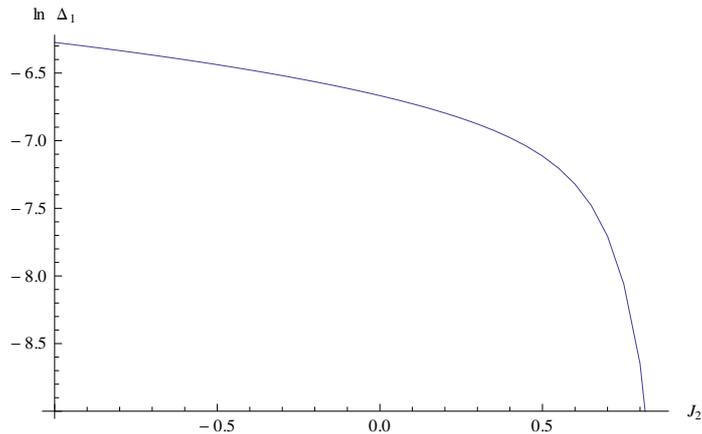}
	\caption{The decreasing behavior of $\ln\Delta_I$ as a function of $J_2$ for $T=0.15$ (right).}
	\label{lngap}	
\end{figure}

\section{Conclusions}
Using the Schwinger boson formalism we have studied the bilinear biquadratic Heisenberg model at
zero and low temperatures. We have shown that, inside the considered limits of $J_2$ (the biquadratic
coupling), the ground state at zero temperature remains ordered. Considering the boson condensation,
we have shown that the degree of  order in a honeycomb lattice is between $58\%$ (for $J_2=-1$) and
$78\%$ (for $J_2=1$), which is weaker than the $81\%$ observed for the square lattice, as expected.
Therefore, even with a smaller coordination number ($z=3$), the quantum fluctuations
in the honeycomb lattice are not sufficient to create a two-dimensional spin liquid state.
Our approach is not appropriate for $|J_2|>1$; however, the asymptotic behavior of the negative values
of $J_2$ (Figure (\ref{density})) suggests the absence of a disordered phase in
the limit $J_2\ll -1$ while for $J_2>1$ the system seems to be strongly ordered.
We have also shown that the ordered ground state exists for all physical spin values.
The superior value of spin for a phase with $\langle m\rangle=0$ is around $0.42$ and it
occurs when $J_2=-1$. Above the ground state, the low energy excitations are
massless Goldstone modes with relativist dispersion relation, since
there is a spontaneous broken symmetry. The spin-wave velocities decrease almost
linearly as a function of $J_2$ (Figure (\ref{spinwave})) and $c_2$ vanishes when
$J_2\approx0.65$ ( $c_1$ vanishes for $J_2>1$, outside the limit considered).
Analyzing the condensate density graphic, one can see that the slow spin-wave velocity corresponds
to a more ordered system (higher condensation). Although the Schwinger formalism is not the
best way to treat finite temperatures, we have found consistent results.
At low temperatures, the ground state is disordered and the excitations have a gap
that increases with the temperature ($\Delta\propto Te^{-\kappa T}$) as
dictated by the Mermin-Wagner theorem.
\label{conclusions}

\section*{Acknowledgment}
The authors would like to thank A. S. T. Pires for comments and suggested articles.
This work was supported by Funda\c c\~ao de Amparo \`a Pesquisa do estado de Minas Gerais
(FAPEMIG) and Conselho Nacional de Desenvolvimento Científico e Tecnológico (CNPq), Brazil.

\end{document}